\documentclass[pdflatex,sn-nature]{sn-jnl}

\usepackage{graphicx}
\usepackage{amsmath,amssymb,amsfonts}
\usepackage{amsthm}
\usepackage[title]{appendix}
\usepackage{xcolor}
\usepackage{textcomp}
\usepackage{booktabs}
\usepackage{multirow}
\usepackage{manyfoot}
\usepackage{comment}
\usepackage{gensymb}

\newcommand{\ket}[1]{\left|#1\right\rangle}
\newcommand{\bra}[1]{\left\langle #1\right|}

\theoremstyle{thmstyleone}%
\theoremstyle{thmstyletwo}%

\theoremstyle{thmstylethree}%

\begin{document}

\title[Article Title]{Emergence of Macroscopic Quantum Order via Translational Zero Modes}



\author*{Kenan Gundogdu} \email{kgundog@ncsu.edu}

\affil{\orgdiv{Department of Physics and Astronomy, Organic and Carbon Electronics Laboratories (ORaCEL)}, \orgname{North Carolina State University}, \orgaddress{\city{Raleigh}, \postcode{27695-7313}, \state{NC}, \country{USA}}}


\abstract{
\begin{quote}
Macroscopic quantum coherence in solids, such as in superfluids, superconductors, and condensates, is generally limited to low temperatures because order forms within a fixed excitation spectrum whose competing states become thermally populated as temperature rises. Here we show that strong coupling between electronic excitations and a deformable lattice enables a different route. Above a critical density, this coupling nucleates self-generated confining potentials that trap the very excitations generating them. Unlike rigid external traps, these potentials can translate through the host lattice without changing their internal structure, defining a translational zero mode. Coupling to this zero mode provides a shared dynamical coordinate that lowers and isolates a single collective many-body configuration, opening a density-dependent gap that suppresses thermal occupation of competing states and supports off-diagonal long-range order at elevated temperatures. As a concrete realization, we identify high-temperature superfluorescence in lead–halide perovskites as the radiative instability of this zero-mode-dressed ordered excitonic state. More broadly, this establishes a general route to macroscopic quantum order: not cooling within a fixed spectrum, nor pairing instabilities, but a self-generated mobile confining structure whose translational zero mode reconstructs the many-body spectrum to protect coherence.
\end{quote}
}

\maketitle

\section*{Main}

The transition to a macroscopic quantum state, whether through Bose--Einstein condensation or a BCS-like pairing instability, is traditionally understood to occur within a pre-existing excitation spectrum~\cite{Leggett2001BECBCS,Cooper1956,BCS1957}. In Bose--Einstein condensation, coherence appears when thermal occupation of excited states can no longer accommodate the particle density, forcing macroscopic occupation of the lowest available state. In BCS-like systems, pairing develops from an already-defined electronic spectrum and opens a gap in the resulting quasiparticle excitations. In both cases, the lattice, trap, or band structure provides the background in which macroscopic order forms.

\begin{figure*}[t]
    \centering
    \includegraphics[width=\textwidth]{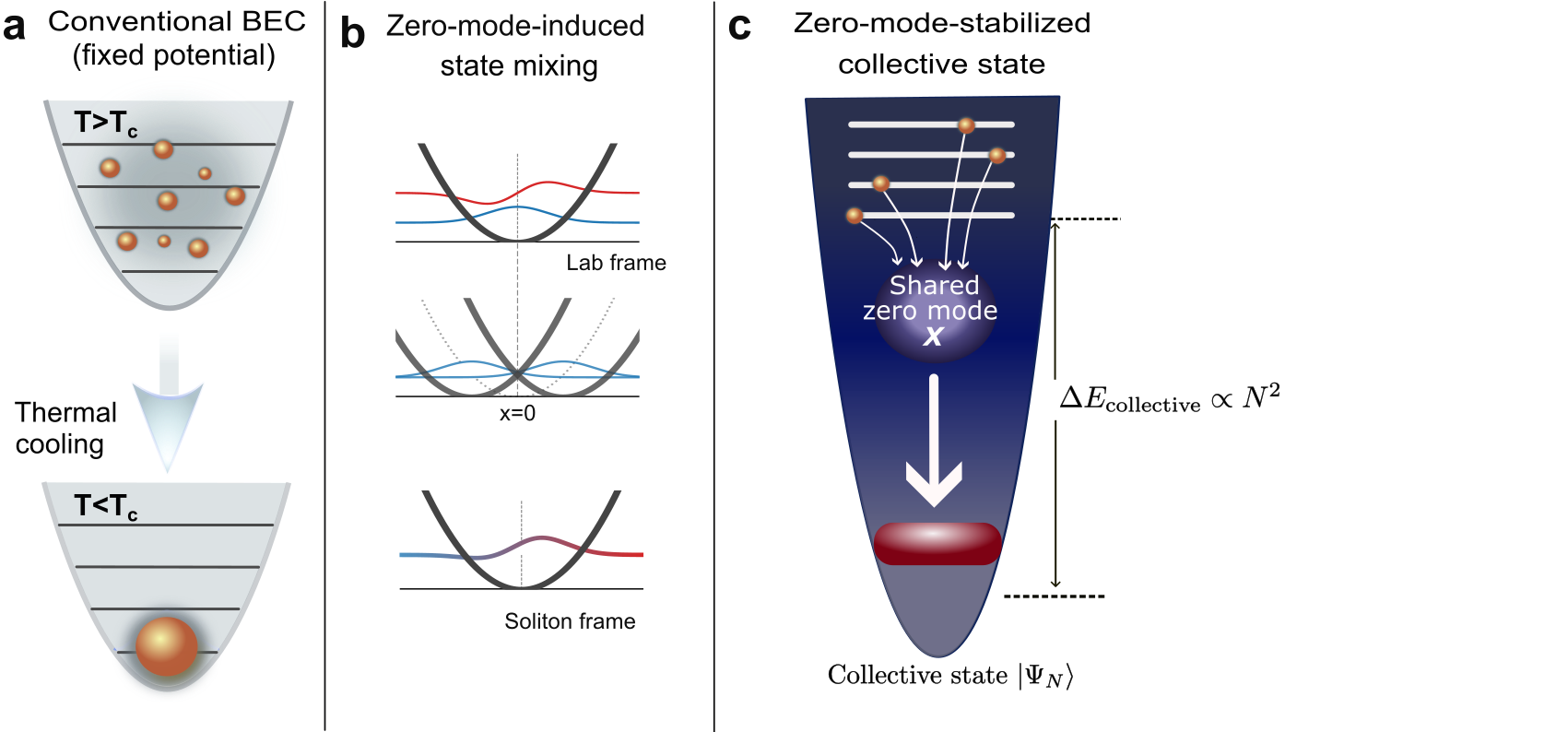}
   \caption{\textbf{From fixed-spectrum condensation to zero-mode-stabilized collective order.}
\textbf{(a)} In conventional condensation, particles occupy the lowest level of a fixed excitation spectrum only after thermal occupation of higher-energy states is reduced below a critical temperature.
\textbf{(b)} In a solitonic confining potential, fluctuations of the translational zero mode shift the potential relative to the laboratory frame. In the soliton frame, this motion coherently mixes the defect-bound excitonic states.
\textbf{(c)} In the many-body regime, all confined excitons couple to the same shared zero-mode coordinate. This collective structural channel drives a single many-body state $|\Psi_N\rangle$ to lower energy, with $\Delta E_{\mathrm{collective}}\propto N^2$, thereby isolating it from the surrounding excitation manifold.}
    \label{fig:figure1}
\end{figure*}

Here, we show that soft crystalline systems offer a fundamentally distinct route. Strong coupling between electronic excitations and a deformable lattice can self-generate a confining potential for the excitations themselves. Because this potential is embedded in an extended host, it translates continuously without changing its internal structure, giving rise to a translational zero mode that acts not as a passive environmental fluctuation but as a shared collective coordinate. Figure~\ref{fig:figure1} illustrates the consequence. A conventional fixed potential (Fig.~\ref{fig:figure1}a) requires thermal depletion of excited levels for macroscopic occupation. By contrast, the mobile self-generated potential (Fig.~\ref{fig:figure1}b) couples the localized electronic states defined relative to it through this shared coordinate. Because all excitations are bound to the same movable structure, the resulting state mixing is collective, producing a cooperative interaction whose leading contribution scales as $N^2$. At high excitation density, this interaction lowers and isolates a single zero-mode-dressed many-body configuration from the surrounding excitation manifold (Fig.~\ref{fig:figure1}c), opening a density-dependent protection gap that suppresses incoherent thermal occupation and stabilizes quantum coherence at elevated temperatures.

This mechanism is motivated by recent observations of high-temperature superfluorescence in lead--halide perovskites~\cite{biliroglu2022room,findik2021high,alanazi2025rtcqe,Tang2025,kobiyama2026transition}. These experiments showed that collective emission, normally suppressed in solids by rapid dephasing, can persist to unexpectedly high temperatures in a soft semiconductor. Subsequent work linked this behavior to a lattice instability in which photoexcited carriers generate solitonic distortions that confine the emitting excitations~\cite{soliton,biliroglu2022room}. The emerging picture is that superfluorescence arises from excitons confined within a self-generated solitonic potential.

This observation raises a sharper question. Is high-temperature superfluorescence merely the result of reduced dephasing, which gives dipoles enough time to synchronize radiatively through the electromagnetic field? Or does the self-generated solitonic landscape play a more fundamental role, reorganizing the excitonic spectrum so that coherence is encoded in the many-body eigenstate before emission? The distinction is essential. In the first case, the lattice protects an otherwise conventional superfluorescent process. In the second, collective emission becomes the optical signature of a macroscopic quantum state stabilized by the deformable solid itself.

Here we develop this second possibility. In this view, the lattice is not a passive source of dephasing, nor merely a medium that extends dipole coherence times; it actively reconstructs the excitation spectrum and stabilizes macroscopic quantum order. Lead--halide perovskites provide the motivating platform, but the principle is more general: deformable quantum materials can, in principle, protect coherence by generating the confining structure and collective coordinate on which the coherent state is built.

Our argument proceeds in four steps. First, photoexcited carriers generate a solitonic confining potential that localizes the relevant excitonic manifold. Second, because this potential is mobile, its translational zero mode provides a shared collective coordinate. Third, coupling to this coordinate induces a cooperative all-to-all interaction among the confined excitons. Fourth, this interaction isolates a zero-mode-dressed many-body configuration from thermally accessible competitors, producing a density-dependent protection gap and stabilizing macroscopic quantum order.

\section*{Self-generated confinement in a deformable solid}

\begin{figure*}[t]
    \centering
    \includegraphics[width=\textwidth]{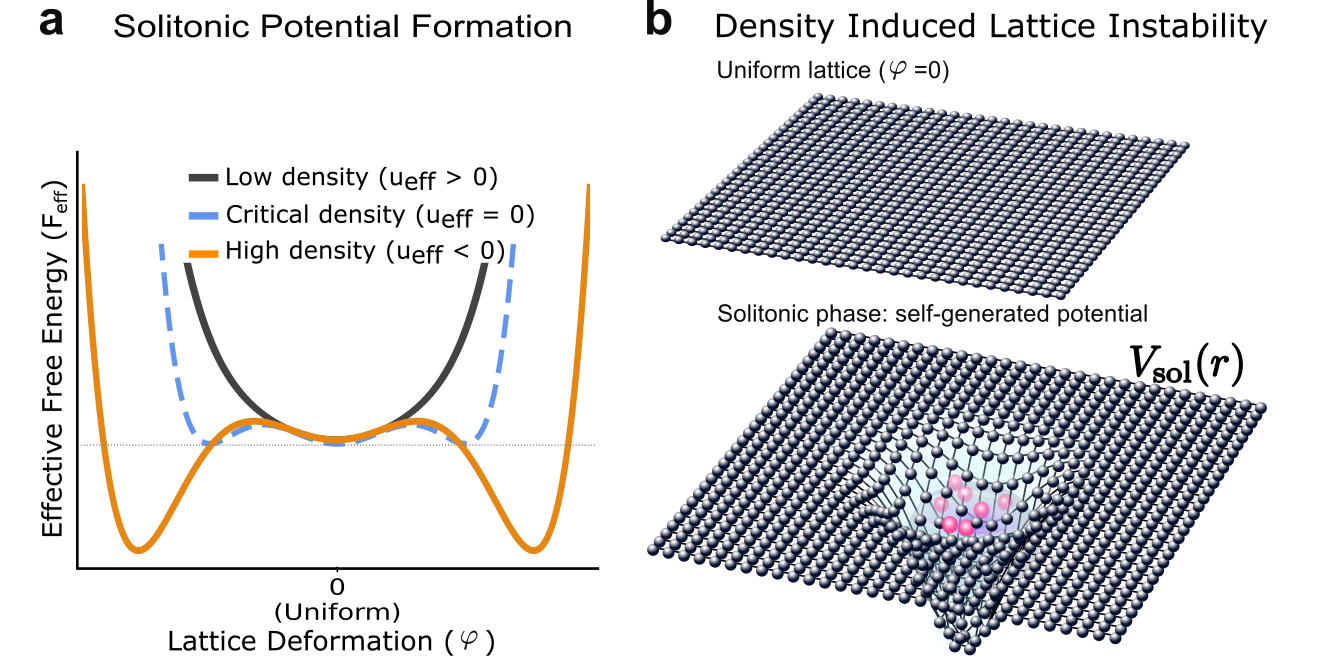}
  \caption{\textbf{Density-induced lattice instability and formation of a self-generated solitonic potential.}
\textbf{(a)} Effective Landau free energy $F_{\mathrm{eff}}$ as a function of the lattice deformation coordinate $\varphi$ for increasing excitation density. At low density, the uniform configuration $\varphi=0$ is the stable minimum. As the density increases, the quadratic exciton--lattice coupling $g_2$ renormalizes the effective quartic coefficient $u_{\mathrm{eff}}$, causing the free-energy landscape to develop finite-deformation minima. At sufficiently high density, a finite-$\varphi$ configuration becomes energetically favored, indicating a thermodynamic instability of the uniform lattice.
\textbf{(b)} Real-space representation of the transition. In the uniform phase, the lattice remains undeformed and no persistent confining potential exists for the excitons. In the solitonic phase, a localized lattice deformation nucleates and creates a self-generated electronic potential $V_{\mathrm{sol}}(\mathbf{r})$. Because the deformation forms within an otherwise uniform host, its absolute position is nearly degenerate, making the potential mobile and giving rise to a translational zero mode.}
    \label{fig:figure2}
\end{figure*}

To establish a concrete microscopic realization of this effect, we first show how the uniform, undeformed lattice can become thermodynamically unstable at high excitation density toward the nucleation of a localized lattice deformation. We consider a minimal microscopic model in which the excitonic density operator, $\hat n(\mathbf{r})$, is coupled to a deformable lattice field, $\varphi(\mathbf{r})$. The total Hamiltonian is written as
\begin{equation}
H = H_{\mathrm{ex}} + H_{\mathrm{lat}} + H_{\mathrm{int}},
\end{equation}
where $H_{\mathrm{ex}}$ describes the excitonic degrees of freedom, $H_{\mathrm{lat}}$ the deformable lattice, and $H_{\mathrm{int}}$ their coupling. The microscopic driving force for the lattice instability is captured by the interaction term
\begin{equation}
H_{\mathrm{int}} = \int d\mathbf{r} \left[g_1 \varphi(\mathbf{r})\hat n(\mathbf{r})+g_2 \varphi^2(\mathbf{r})\hat n(\mathbf{r})\right].
\end{equation}
Here, the linear coupling $g_1$ produces ordinary polaronic shifts and provides a channel for dephasing. By contrast, the formation of the solitonic defect arises primarily from the quadratic exciton--lattice coupling, $g_2$. At low excitation density, the lattice remains in a uniform configuration with $\varphi=0$ (Fig.~\ref{fig:figure2}b), and its stability is described by a symmetric Landau free-energy potential (Fig.~\ref{fig:figure2}a). As the excitation density increases, exciton--lattice interactions renormalize the effective Landau free energy,
\begin{equation}
F_{\mathrm{eff}}[\varphi]
=
r_{\mathrm{eff}}\varphi^2
+
u_{\mathrm{eff}}\varphi^4
+
v_{\mathrm{eff}}\varphi^6
+\cdots .
\end{equation}

As detailed in Supplementary Note1, integrating out local excitonic density fluctuations gives
\begin{equation}
u_{\mathrm{eff}}
=
u_{\mathrm{lat}}
-
2\beta g_2^2 \mathrm{Var}_0(\hat n),
\end{equation} 
where $\beta=1/k_BT$ is the inverse temperature,  $\hat n$ denotes the coarse-grained exciton density in a local region, and $\mathrm{Var}_0(\hat n)$ measures the susceptibility of the exciton population to forming density inhomogeneity around the undeformed state.

This expression shows that the quadratic exciton--lattice coupling softens the lattice anharmonicity. When $2\beta g_2^2 \mathrm{Var}_0(\hat n_c) = u_{\mathrm{lat}}$, the effective quartic coefficient changes sign, and the uniform $\varphi=0$ lattice is no longer the globally favored configuration. Instead, the free-energy landscape develops a lower-energy minimum at finite lattice deformation amplitude. This behavior is illustrated schematically in Fig.~\ref{fig:figure2}. In contrast to a conventional continuous soft-mode transition \cite{chaikin1995principles}, the symmetric uniform configuration, $\varphi=0$, can remain a local minimum while ceasing to be the global thermodynamic minimum. The effective free-energy landscape is therefore reconfigured into a metastable form: the uniform state is separated by a finite barrier from a lower-energy finite-$\varphi$ configuration. Once the excitation density is sufficiently high, the system crosses this barrier and nucleates localized solitonic domains. This coarse-grained analysis establishes the instability; the nucleated state is a localized soliton of finite size, derived variationally and observed directly in Ref.~\cite{soliton}.

\begin{figure*}[b]
    \centering
    \includegraphics[width=0.8\textwidth]{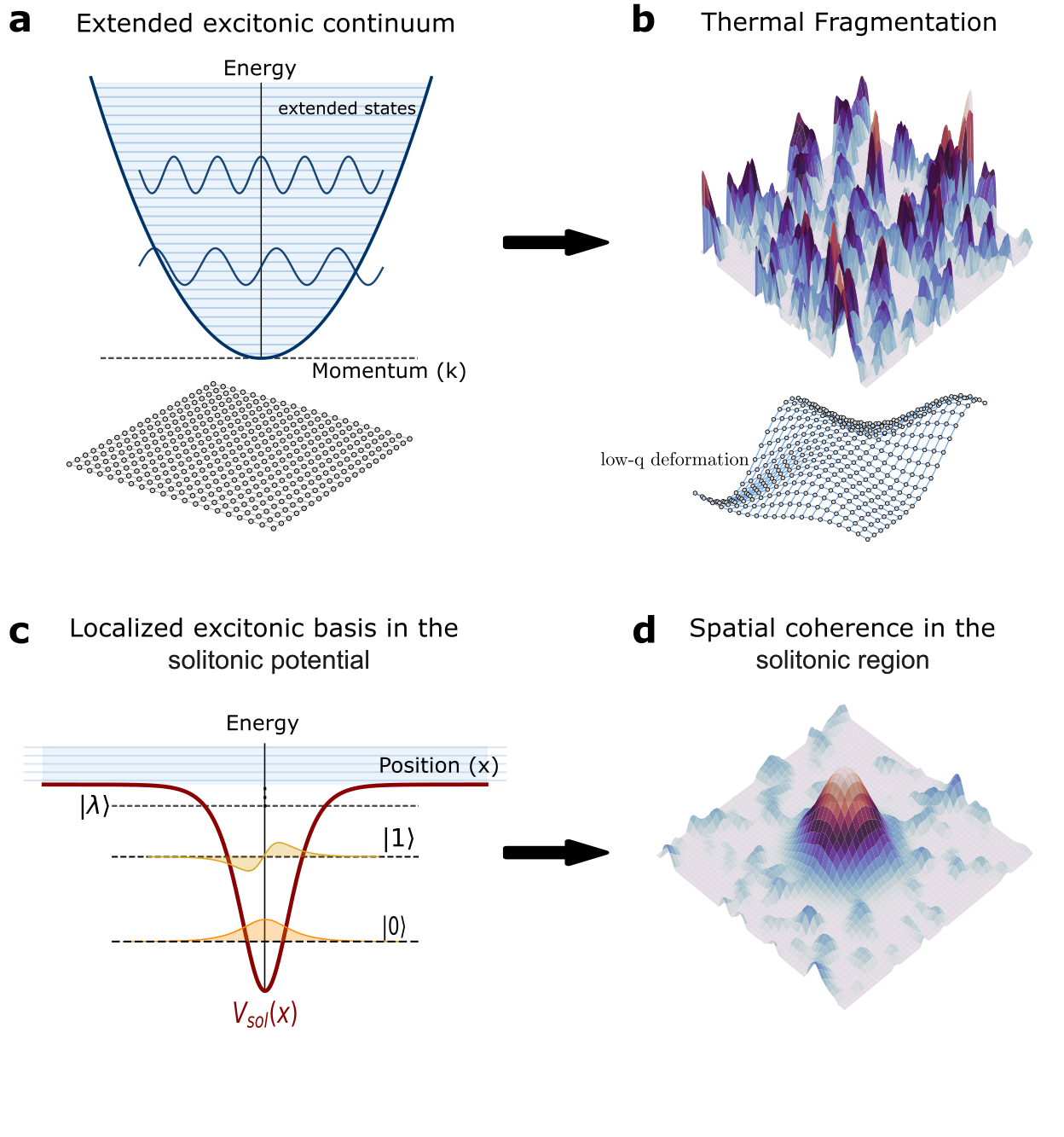}
 \caption{\textbf{Impact of the solitonic transition on excitonic spatial coherence and potential landscapes.}
\textbf{(a)} In the homogeneous regime, the uniform crystal lattice defines an unconfined continuous spectrum, supporting extended excitonic eigenstates approximated here as plane waves.
\textbf{(b)} Extended excitons interact with a gapless continuum of traveling thermal phonons. These low-energy fluctuations generate random lattice distortions that modulate the potential landscape, dephase the excitonic wavefunctions, and distribute the population over many spatially incoherent configurations.
\textbf{(c)} The solitonic transition changes the electronic geometry. A localized structural deformation generates a confining potential, $V_{\mathrm{sol}}(\mathbf{r})$, which mixes the delocalized plane-wave basis and projects the excitonic spectrum into a discrete manifold of localized bound states, denoted by the $|\lambda\rangle$ basis.
\textbf{(d)} Once the spectrum becomes discrete and the relevant lattice fluctuations are localized to the solitonic region, the long-wavelength phonon channels responsible for spatial dephasing of extended states are strongly suppressed.}
    \label{fig:figure3}
\end{figure*}

The structural transition fundamentally redefines the electronic environment. Figure~\ref{fig:figure3} contrasts the two phases. In the uniform crystal, the lattice supports a continuous band of extended excitonic eigenstates (Fig.~\ref{fig:figure3}a). These extended states couple directly to a gapless continuum of propagating thermal phonons (Fig.~\ref{fig:figure3}b), whose low-energy fluctuations provide many scattering and dephasing channels. As a result, the excitonic population is distributed over many extended states with rapidly randomized relative phases, producing a fragmented, incoherent ensemble.

In the solitonic phase, the lattice dynamics are organized around the local deformation. Decomposing the lattice field as $\varphi(\mathbf{r}) = \varphi_{\text{sol}}(\mathbf{r}) + \delta\varphi(\mathbf{r})$, the static component generates, to leading order, an effective confining potential
\begin{equation}
V_{\text{sol}}(\mathbf{r}) = g_1 \varphi_{\text{sol}}(\mathbf{r}) + g_2 \varphi_{\text{sol}}^2(\mathbf{r}),
\end{equation}
whose single-particle excitonic states are obtained by solving the Schr\"{o}dinger equation in $V_{\text{sol}}(\mathbf{r})$ (Supplementary Note~2). This localized potential converts the extended excitonic spectrum into a discrete set of localized bound states, denoted by the $|\lambda\rangle$ basis (Fig.~\ref{fig:figure3}c).

The localization also changes how the excitons couple to lattice fluctuations. Once the spectrum is discrete and the relevant fluctuations are confined to the solitonic region, the long-wavelength phonon channels that dephased and fragmented the extended states no longer act efficiently in real space (Fig.~\ref{fig:figure3}d). Their primary residual effect is to induce mixing and thermal redistribution within the localized bound-state manifold (Supplementary Note~2).

\section*{The translational zero mode as a shared coordinate}

Up to this point, the solitonic trap behaves like any quantum-confined system: internal states $|0\rangle$ and $|\lambda\rangle$ set by a confining potential, with phonons acting only as a bath that redistributes population among them. The decisive difference is that this trap can move. Because the deformation forms within a uniform host, it can translate without changing its internal structure, so its position is a soft collective coordinate, a translational zero mode, that the internal states $|0\rangle$ and $|\lambda\rangle$ do not capture. In what follows we show that many excitons can use this shared coordinate to lower their total energy by condensing into a single collective orbital. We establish this in two stages. First, we show that fluctuations along the zero mode mix the confined electronic states through an unusually strong coupling matrix element. Second, we show that in the many-body regime this single coordinate dominates over all other lattice fluctuations, acting as one coherent channel rather than an incoherent thermal bath.

As the soliton slides along its translational collective coordinate, the confining potential shifts relative to a reference configuration. For a small spatial displacement $\delta \mathbf{R}$, the induced perturbation is
\begin{equation}
\delta V(\mathbf{r})
=
-\delta \mathbf{R}\cdot \nabla V_{\mathrm{sol}}(\mathbf{r}) .
\end{equation}
The coupling between the localized ground state $|0\rangle$ and an excited bound state $|\lambda\rangle$ is therefore governed by the matrix element
\begin{equation}
    A_{0\lambda}
    =
    \langle 0|\nabla V_{\mathrm{sol}}(\mathbf{r})|\lambda\rangle .
    \label{eq:zero-mode-matrix-element}
\end{equation}
The central feature of this coupling is its intrinsic strength. Because the perturbation operator $\nabla V_{\mathrm{sol}}(\mathbf{r})$ is the spatial derivative of the same potential that defines the confined states, it is geometrically matched to their wavefunctions, sharing the same characteristic length scales and symmetry structure. As shown analytically in Supplementary Note~3.2, for a one-dimensional harmonic confining potential this matrix element scales as $A_{01}\sim \Delta E_{10}/\ell_{\mathrm{ex}}$, where $\Delta E_{10}=E_1-E_0$ and $\ell_{\mathrm{ex}}$ is the spatial extent of the localized excitonic wavefunction. Thus a displacement over the exciton size produces a mixing energy comparable to the internal level spacing, demonstrating that the zero mode provides an unusually efficient coupling between the bound excitonic states.

For a single particle, a rigid translation of the solitonic potential leaves the exact eigenvalue spectrum unchanged when the full translated basis is treated consistently. The zero mode becomes consequential only when many excitons occupy the same mobile confining landscape. To determine its role there, and why it stands out among all the lattice degrees of freedom, we integrate out the lattice to obtain an exciton-only effective Hamiltonian (Supplementary Note~4):
\begin{equation}
H_{\mathrm{eff}}
=
H_{\mathrm{ex}}
-
\sum_{\alpha}
\frac{1}{2\kappa_{\alpha}}
\hat{A}_{\alpha}^{2}.
\label{eq:integrated-out-lattice-hamiltonian}
\end{equation}
Here, $H_{\mathrm{ex}}$ is the bare many-body excitonic Hamiltonian, $\alpha$ indexes the available vibrational modes, $\kappa_{\alpha}$ is the effective dynamic stiffness of mode $\alpha$, and $\hat{A}_{\alpha}$ is the corresponding exciton--lattice coupling operator. The coupling strength and stiffness of each mode therefore determine its relative impact on the excitonic spectrum.

Equation~\eqref{eq:integrated-out-lattice-hamiltonian} makes clear why the translational zero mode dominates. Because the solitonic defect resides in an otherwise uniform host, this mode is anomalously soft, with an effective stiffness $\kappa$ strongly reduced relative to ordinary vibrational modes; and, as shown above, it also carries a large matrix element [Eq.~\eqref{eq:zero-mode-matrix-element}]. Having both a small stiffness and a large coupling, it provides the dominant coherent many-body channel inside the solitonic domain. In the following we show that coupling to this mode drives the excitons into a collective orbital whose energy is lowered by an amount scaling as $N^2$.

To demonstrate this, we consider the operator through which all $N$ excitons couple to the zero mode, the sum of their single-particle couplings,
\begin{equation}
\hat{A}
=
\sum_{i=1}^{N}
\nabla_i V_{\mathrm{sol}}(\mathbf{r}_i)
\equiv
\sum_{i=1}^{N}
\hat{A}_i .
\label{eq:collective-zero-mode-operator}
\end{equation}
This operator enters the effective Hamiltonian [Eq.~\eqref{eq:integrated-out-lattice-hamiltonian}] through the squared term $\hat{A}^{2}$. Expanding it over the $N$ excitons separates the self and cross contributions:
\begin{equation}
\hat{A}^{2}
=
\left(
\sum_{i=1}^{N}
\hat{A}_{i}
\right)
\cdot
\left(
\sum_{j=1}^{N}
\hat{A}_{j}
\right)
=
\underbrace{
\sum_{i=1}^{N}
\hat{A}_{i}^{2}
}_{\mathrm{Self\ terms}}
+
\underbrace{
\sum_{i\neq j}
\hat{A}_{i}\cdot \hat{A}_{j}
}_{\mathrm{Cross\ terms}} .
\end{equation}
The self terms describe how each exciton individually responds to a rigid displacement of the solitonic trap. The cross terms describe how two different excitons respond to the \emph{same} displacement of the shared confining potential. The fate of these cross terms depends on the relative phases of the excitons. If the excitons occupy independent states with random relative phases, their responses to the zero mode interfere destructively, and the cross-term sum averages to zero. If instead the excitons share a common orbital with locked phases, their responses interfere constructively and add up, producing a collective energy reduction that scales as $N(N-1)$. Energy minimization therefore drives the excitons into a single phase-locked, zero-mode-dressed collective orbital.

Within a mean-field treatment (Supplementary Note 4.2) of the shared zero-mode coordinate, the energetic consequence of forming this orbital is a collective lowering of the many-body configuration, with characteristic scale:
\begin{equation}
\Delta E_{\mathrm{collective}}
\sim
-
\frac{N^2 |A_{0\lambda}|^2}{2\kappa}.
\label{eq:collective-energy-lowering}
\end{equation}
The resulting energy lowering per particle grows linearly with density, thereby opening a large collective gap:
\begin{equation}
    \Delta_{\text{gap}} \propto \frac{N |A_{0\lambda}|^2}{\kappa}.
    \label{density_dependent_gap}
\end{equation}
Unlike rigid, externally engineered traps, whose confinement scale is fixed, this density-driven gap strengthens as the collective state builds up and thereby suppresses scattering into thermally accessible configurations.

The other vibrational modes are not removed by this dominance. Having finite stiffness, they contribute with smaller weight and act as competing, incoherent channels, driving scattering, dephasing, and wavefunction renormalization. In the solitonic phase they primarily redistribute population within the localized bound-state manifold. These modes constitute the thermal background against which the collective gap must compete.

\section*{Zero-mode enhanced condensation and off-diagonal long-range order}

The appearance of this macroscopic, density-dependent gap is the central thermodynamic consequence of the zero-mode interaction. It does more than simply lower the energy of a collective state: it fundamentally alters the thermal capacity of the excited manifold itself. In a conventional condensate, macroscopic occupation occurs only when the thermal population in the fixed excited-state manifold becomes saturated at low temperatures. Here, the thermodynamic mechanism is inverted. As the particle density increases, the cooperative zero-mode interaction drives the collective state progressively deeper below the thermally accessible manifold. Consequently, the exact same particle population that builds the coherent state simultaneously suppresses the phase space available for incoherent excitations.

\begin{figure*}[t]
    \centering
    \includegraphics[width=\textwidth]{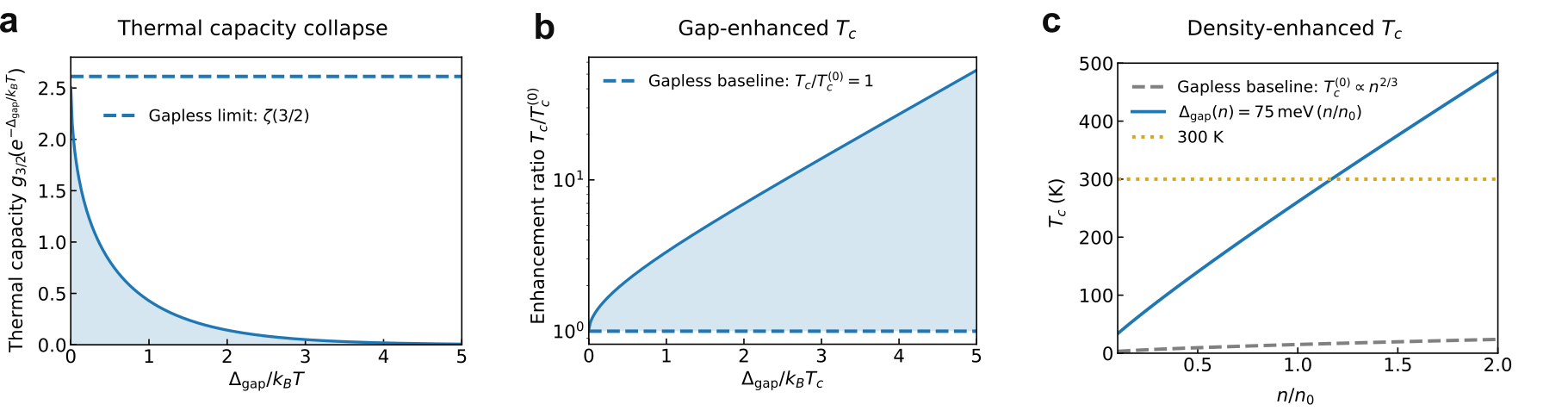}
   \caption{\textbf{Thermal-capacity suppression by a density-dependent collective gap.}
\textbf{a,} A finite $\Delta_{\rm gap}$ reduces the Bose phase-space factor from the gapless value $g_{3/2}(1)=\zeta(3/2)$ to $g_{3/2}\!\left(e^{-\Delta_{\rm gap}/k_B T}\right)$, suppressing the thermal capacity of the excited manifold. 
\textbf{b,} This suppression increases the critical temperature relative to the gapless baseline $T_c^{(0)}$, shown by the dashed line. 
\textbf{c,} For a representative cooperative gap $\Delta_{\rm gap}(n)=75~\mathrm{meV}\,(n/n_0)$, the self-consistent $T_c$ rises more rapidly than the conventional $T_c^{(0)}\propto n^{2/3}$ scaling. The value of $\Delta_{\rm gap}$ is illustrative; the absolute scale is set by material parameters. At high density this rise is ultimately bounded by the soliton-formation condition, which is not included here. The dotted line marks 300~K.}
    \label{fig:figure4}
\end{figure*}

To formalize this density-driven condensation, we calculate the thermal capacity of the excited manifold, defined as the maximum exciton density that can be accommodated outside the lowest collective mode at a given temperature (Fig.~\ref{fig:figure4}). When the total exciton density exceeds this capacity, the excess population must accumulate in the lowest collective mode. For a conventional trapped condensate in a static potential, where no collective gap is present ($\Delta_{\rm gap}=0$), this condition yields the standard critical temperature $T_c^{(0)}$, governed by the Riemann zeta function $\zeta(3/2)$ and the algebraic scaling $T_c^{(0)}\propto n^{2/3}$~\cite{pethick2008bose}. In our structurally mediated system, however, the collective gap reduces the phase-space capacity of the excited manifold. As derived in Supplementary Note~6, this raises the critical temperature according to
\begin{equation}
    \frac{T_c}{T_c^{(0)}} =
    \left[
    \frac{\zeta(3/2)}
    {g_{3/2}\left(e^{-\Delta_{\text{gap}}/k_B T_c}\right)}
    \right]^{2/3},
\end{equation}
where $g_{3/2}(z)$ is the Bose function, or polylogarithm, evaluated at the effective fugacity set by the zero-mode gap.

Figure~\ref{fig:figure4}a shows that a finite gap reduces the Bose phase-space factor from its gapless value $g_{3/2}(1)=\zeta(3/2)$ to $g_{3/2}\!\left(e^{-\Delta_{\mathrm{gap}}/k_B T}\right)$, suppressing the thermal capacity of the excited manifold so that a given exciton density saturates it at progressively higher temperatures (Fig.~\ref{fig:figure4}b). The essential feature is that this gap is not externally imposed but generated cooperatively by the condensing population itself, and therefore grows with density, $\Delta_{\mathrm{gap}}\propto N$. This creates a nonlinear feedback loop [Eq.~\eqref{density_dependent_gap}]: higher density deepens the collective minimum, which further reduces the excited-state capacity and drives still more population into the collective mode. The statistical analysis is the same as for conventional condensation, but because the gap is self-generated rather than fixed, the spectrum and its occupation must be solved self-consistently. As Fig.~\ref{fig:figure4}c shows, this feedback drives $T_c$ above the conventional $n^{2/3}$ scaling.

This collapse of the excited-state capacity is the thermodynamic manifestation of the rank-one structure of the zero-mode interaction. Because it is the square of a single collective operator, the interaction is separable: rather than scattering excitons across a broad manifold, it projects them into one dominant channel. In Supplementary Note~6 we minimize the many-body energy in this channel variationally, obtain the explicit zero-mode-dressed collective orbital, and show that the resulting one-body density matrix develops a single leading eigenvalue of order $N$ while all others remain of order unity. This is the Penrose--Onsager criterion for off-diagonal long-range order (ODLRO), the defining hallmark of macroscopic quantum coherence in superfluids and superconductors~\cite{penrose1956bose,yang1962concept}. Crucially, whereas those conventional condensates achieve ODLRO by cooling within a fixed excitation spectrum, here it emerges through density-dependent dynamical reconstruction of the spectrum.

\section*{Experimental signatures and broader implications}

Realization of this mechanism fundamentally depends on the formation of the underlying confining solitonic potential (Fig.~\ref{fig:figure2}). Thus, achieving coherence at elevated temperatures demands host materials with pronounced nonlinear lattice flexibility, capable of supporting stable, localized distortions under thermal conditions.  These criteria are naturally satisfied by soft materials with strong electron--lattice coupling, including two-dimensional transition metal dichalcogenides~\cite{weston2020atomic} and perovskites~\cite{wright2016electron,yazdani2024coupling,guzelturk2021visualization}. Depending on the material system and the nature of the collective state, this mechanism should lead to distinct experimentally testable signatures. In lead--halide perovskites, the observed high-temperature superfluorescence can be understood as the radiative instability of the macroscopically coherent excitonic state stabilized within the solitonic trap.  Such collective coherence at elevated temperatures is difficult to reconcile with conventional condensation. Within a fixed excitation spectrum, room-temperature macroscopic coherence is accessible only when the effective mass is extremely small \cite{plumhof2014room,kasprzak2006bose}, as in exciton--polariton condensates, where it is moreover sustained under continuous driving rather than in equilibrium. The persistence of cooperative emission to high temperatures in these heavy-mass excitonic systems is therefore anomalous within the standard paradigm, yet follows naturally if the coherence is protected by a self-generated structural gap. We stress that this provides a consistent microscopic origin for an already-observed phenomenon rather than a proof of the mechanism; the discriminating tests are described below. 

A key ingredient of this interpretation is that the excitonic states confined within the solitonic manifold preserve mutual coherence and do not fragment spatially. This yields a direct optical test. Because all excitons in the solitonic manifold share a single phase-locked collective orbital, their transition dipoles add coherently, so the mechanism predicts that resonant absorption into the defect-bound manifold scales with the number of condensed excitons rather than additively across independent states, a collectively enhanced oscillator strength that is the absorptive counterpart of the cooperative emission seen in superfluorescence~\cite{boehme2025single}. Crucially, this enhancement appears only if the solitonic deformation is prepared before the excitonic manifold is populated; its presence therefore discriminates our mechanism from a conventional dephasing-limited picture, in which no such cooperative absorption is expected. A dedicated experimental test of this prediction will be reported separately.

A second signature concerns how the onset temperature for collective emission scales with excitation density. Because the protective gap is generated by the condensed population itself, the onset temperature should rise faster with density than the $n^{2/3}$ law of condensation in a fixed spectrum (Fig.~\ref{fig:figure4}c). This rise is bounded by soliton formation: the underlying structural instability is thermally suppressed (Supplementary Note~1), defining a density-dependent ceiling above which no trap nucleates. The onset temperature is therefore predicted to climb steeply at low and intermediate density and then saturate as it meets this ceiling, a combined behavior with which existing high-temperature superfluorescence data are qualitatively consistent~\cite{biliroglu2022room,soliton}.

A third signature appears in the macroscopic polarization, which reports on the collective coherent order parameter through the superfluorescent instability. Because the density-dependent gap $\Delta_{\rm gap}$ sets the scale against $k_B T$, the fluctuation (Ginzburg) regime should acquire a marked temperature and density dependence rather than a fixed width. The order-parameter exponent is likewise predicted to depart from standard values (mean-field $1/2$, three-dimensional XY $\approx 0.35$)\cite{campostrini2001critical}, because the coherent amplitude is limited by boundary noise penetration and therefore improves with the surface-to-volume ratio of the solitonic domain. The nucleation dynamics underlying the onset-temperature scaling and the detailed analysis of this Ginzburg regime are presented in a separate study.

More broadly, the mechanism is not tied uniquely to excitons. The same structure should arise whenever an electronic manifold is confined by an emergent deformation whose position, phase, or registry remains a soft collective coordinate, which then plays the role of the soliton translation mode and reorganizes the effective many-body spectrum. By providing a shared dynamical coordinate, such a mode may also mediate or protect the correlations needed to drive a pairing instability, so the principle extends beyond bosonic condensation to fermionic order. Candidate realizations include moir\'e systems with soft phason or registry modes and correlated excitonic phases~\cite{Moire_1_maity2023electrons,weston2020atomic,Xiong2023CorrelatedExcitonInsulator,Xie2023NematicExcitonicInsulator}, twisted van der Waals systems with structural relaxation~\cite{Cantele2020StructuralRelaxationTBG}, and stripe phases in correlated systems~\cite{kivelson1998electronic,tranquada1995evidence}, with the geometry of the confining structure setting the dimensionality of the resulting correlations. These results challenge the assumption that macroscopic quantum order must emerge within a fixed excitation spectrum, with the lattice acting only as a static background or thermal bath. Instead, strong coupling to a deformable host can reconstruct the excitation spectrum itself, with the medium actively constructing and protecting the collective state. Because zero modes act as shared dynamical coordinates, they imprint their symmetry onto the effective many-body interaction, enabling chiral, valley-dependent, or topological functionalities with no precedent in fixed-spectrum materials, and establishing soft-coordinate engineering as a route to coherent phases in deformable quantum systems.

\bibliography{sn-bibliography}

\section*{Acknowledgements}
The author acknowledges discussions with M. Biliroglu, M. Unsal and D. Aspnes. 

\section*{Funding}
This material is based upon work supported by the U.S. Department of Energy, Office of Science, Basic Energy Sciences, under Award No. DE-SC0024396.

\section*{Competing Interests}
The author declares that they have no competing interests.

\section*{Materials and Correspondence}
All the simulation plots within this paper are available from the corresponding author upon reasonable request.

\clearpage
\setcounter{equation}{0}
\renewcommand{\theequation}{S\arabic{equation}}

\setcounter{figure}{0}
\renewcommand{\thefigure}{S\arabic{figure}}

\setcounter{table}{0}
\renewcommand{\thetable}{S\arabic{table}}

\renewcommand{\theHequation}{S\arabic{equation}}
\renewcommand{\theHfigure}{S\arabic{figure}}
\renewcommand{\theHtable}{S\arabic{table}}

\section*{Supplementary Information}

\section*{Supplementary Note 1: Fluctuation-Driven Thermodynamic Instability and Solitonic Trap Formation}
In this section we will show that the quadratic term in the exciton lattice interaction Hamiltonian (Eq.1 in main text) leads to a structural instability and the nucleation of solitonic traps. 
We begin with the total Hamiltonian
\begin{equation}
H = H_{\mathrm{ex}} + H_{\mathrm{lat}} + H_{\mathrm{int}}.
\end{equation}

To capture spatial redistribution of excitons, we consider a coarse-grained region of fixed volume $\Omega$, within which the lattice distortion is approximately uniform and equal to $\varphi$. Let $\hat{N}_\Omega$ denote the exciton number operator in this region. The quadratic exciton--lattice coupling then reduces to
\begin{equation}
H_{\mathrm{int}}^{(2)} = g_2 \varphi^2 \hat{N}_\Omega.
\end{equation}

The bare lattice free energy of this region is written in Landau form as
\begin{equation}
F_{\mathrm{lat}}(\varphi)
=
\frac{1}{2} r_{\mathrm{lat}} \varphi^2
+
\frac{1}{4} u_{\mathrm{lat}} \varphi^4
+\cdots.
\end{equation}

At fixed $\varphi$, the excitonic Hamiltonian becomes
\begin{equation}
H_{\mathrm{ex}}(\varphi) = H_0 + g_2 \varphi^2 \hat{N}_\Omega,
\end{equation}
where $H_0$ is the unperturbed excitonic Hamiltonian.

The corresponding partition function is
\begin{equation}
Z_{\mathrm{ex}}(\varphi)
=
\mathrm{Tr}\, e^{-\beta H_{\mathrm{ex}}(\varphi)}
=
\mathrm{Tr}\, e^{-\beta (H_0 + g_2 \varphi^2 \hat{N}_\Omega)}.
\end{equation}

Defining
\begin{equation}
Z_0 = \mathrm{Tr}\, e^{-\beta H_0},
\qquad
\langle \cdots \rangle_0
=
\frac{1}{Z_0}\mathrm{Tr}\left[(\cdots)e^{-\beta H_0}\right],
\end{equation}
we rewrite
\begin{equation}
Z_{\mathrm{ex}}(\varphi)
=
Z_0
\left\langle
e^{-\beta g_2 \varphi^2 \hat{N}_\Omega}
\right\rangle_0.
\end{equation}

The excitonic free energy is
\begin{equation}
F_{\mathrm{ex}}(\varphi)
=
-\frac{1}{\beta} \ln Z_{\mathrm{ex}}(\varphi)
=
F_0
-
\frac{1}{\beta}
\ln
\left\langle
e^{-\beta g_2 \varphi^2 \hat{N}_\Omega}
\right\rangle_0,
\end{equation}
with $F_0 = -\frac{1}{\beta}\ln Z_0$.

Introducing
\begin{equation}
X = -\beta g_2 \varphi^2 \hat{N}_\Omega,
\end{equation}
and expanding to second order using the cumulant expansion,
\begin{equation}
\ln \langle e^X \rangle_0
\approx
\langle X \rangle_0
+
\frac{1}{2}
\left(
\langle X^2 \rangle_0 - \langle X \rangle_0^2
\right),
\end{equation}
we obtain
\begin{equation}
\langle X \rangle_0
=
-\beta g_2 \varphi^2 \langle \hat{N}_\Omega \rangle_0,
\end{equation}
and
\begin{equation}
\langle X^2 \rangle_0 - \langle X \rangle_0^2
=
\beta^2 g_2^2 \varphi^4
\mathrm{Var}_0(\hat{N}_\Omega),
\end{equation}
where
\begin{equation}
\mathrm{Var}_0(\hat{N}_\Omega)
=
\langle \hat{N}_\Omega^2 \rangle_0
-
\langle \hat{N}_\Omega \rangle_0^2.
\end{equation}

Substituting back, the excitonic free energy becomes
\begin{equation}
F_{\mathrm{ex}}(\varphi)
\approx
F_0
+
g_2 \langle \hat{N}_\Omega \rangle_0 \varphi^2
-
\frac{\beta}{2} g_2^2 \mathrm{Var}_0(\hat{N}_\Omega) \varphi^4.
\end{equation}

Adding the lattice contribution,
\begin{equation}
F_{\mathrm{eff}}(\varphi)
=
F_{\mathrm{lat}}(\varphi)
+
F_{\mathrm{ex}}(\varphi),
\end{equation}
we obtain
\begin{equation}
F_{\mathrm{eff}}(\varphi)
=
F_0
+
\frac{1}{2} r_{\mathrm{eff}} \varphi^2
+
\frac{1}{4} u_{\mathrm{eff}} \varphi^4
+\cdots,
\end{equation}
with renormalized coefficients
\begin{equation}
r_{\mathrm{eff}}
=
r_{\mathrm{lat}}
+
2 g_2 \langle \hat{N}_\Omega \rangle_0,
\end{equation}
\begin{equation}
u_{\mathrm{eff}}
=
u_{\mathrm{lat}}
-
2 \beta g_2^2 \mathrm{Var}_0(\hat{N}_\Omega).
\end{equation}

The quartic term becomes unstable when
\begin{equation}
u_{\mathrm{eff}} < 0,
\end{equation}
i.e.,
\begin{equation}
2 \beta g_2^2 \mathrm{Var}_0(\hat{N}_\Omega) > u_{\mathrm{lat}}.
\end{equation}

This demonstrates that fluctuations of exciton number within a coarse-grained region generate a negative correction to the quartic lattice stiffness, enabling a finite-amplitude structural instability even when the total exciton number in the system is fixed.

\section*{Supplementary Note 2: How the Static Soliton Reorganizes the Exciton Spectrum}

Following the structural phase transition described in Supplementary Note 1, the continuous lattice deformation field $\varphi(\mathbf{r})$ acquires a macroscopic, localized static profile. In this section, we quantize the effective Hamiltonian within this solitonic background, derive the bare localized excitonic basis. 
This establishes the static excitonic framework required before introducing the dynamic zero-mode fluctuations in Note 3.

\subsection*{2.1 The Static Soliton Acts as a Local Confining Potential}

We begin by separating the continuous lattice deformation field into the newly formed static solitonic profile, $\varphi_0(x)$, and the dynamic fluctuations around it, $\delta\varphi(\mathbf{r})$:
\begin{equation}
    \varphi(\mathbf{r}) = \varphi_0(\mathbf{r}) + \delta\varphi(\mathbf{r})
\end{equation}
with $\langle \delta\varphi(x) \rangle = 0$. Substituting this decomposition into the interaction Hamiltonian $H_{\text{int}}$, the purely static contribution acting on the excitonic sector is:
\begin{equation}
    H_{\text{static}} = \int d\mathbf{r} \left[ g_1 \varphi_0(\mathbf{r}) + g_2 \varphi_0^2(\mathbf{r}) \right] \hat{n}(\mathbf{r})
    \label{eqHstatic}
\end{equation}
where $\hat{n}(\mathbf{r}) = \psi^\dagger(\mathbf{r})\psi(\mathbf{r})$ is the local exciton density operator. To evaluate how this static deformation reconstructs the electronic Hilbert space, we project the field operator onto the unperturbed, translationally invariant bare excitonic basis $w_n(\mathbf{r})$, such that $\psi(\mathbf{r}) = \sum_n w_n(\mathbf{r}) c_n$. The static interaction Hamiltonian becomes:
\begin{equation}
    H_{\text{static}} = \sum_{nm} J_{nm} c_n^\dagger c_m
\end{equation}
where the bilinear coupling kernel $J_{nm}$ is given by the spatial overlap between the solitonic potential and the unperturbed wavefunctions:
\begin{equation}
    J_{nm} = \int d\mathbf{r} \left[ g_1 \varphi_0(\mathbf{r}) + g_2 \varphi_0^2(\mathbf{r}) \right] w_n^*(\mathbf{r}) w_m(\mathbf{r})
\end{equation}

\subsection*{2.2 The Static Soliton Creates a Local Potential Well}

To elucidate the physical intuition behind the coupling matrix $J_{nm}$, it is useful to transition from the many-body field formulation to the first-quantized framework of a single localized exciton.

When evaluating the total static interaction in Eq.~\ref{eqHstatic}, the density operator is integrated over the entire system to obtain the total static interaction energy. For a single excitation, however, the relevant object is the local potential landscape experienced at position $\mathbf{r}$.

Treating the solitonic deformation $\varphi_0(\mathbf{r})$ as a static background, the single-particle Hamiltonian is

\begin{equation}
H_{\text{single}} = \int d\mathbf{r} \, \psi^\dagger(\mathbf{r})
\underbrace{\Big[ h_0 + g_1 \varphi_0(\mathbf{r}) + g_2 \varphi_0^2(\mathbf{r}) \Big]}_{\hat{H}_{\text{QM}}}
\psi(\mathbf{r}) .
\end{equation}

The bracketed term defines the quantum mechanical Hamiltonian for one exciton,
\begin{equation}
\hat{H}_{\text{QM}} = h_0 + V_{\text{eff}}(\mathbf{r}),
\end{equation}
with
\begin{equation}
V_{\text{eff}}(\mathbf{r}) = g_1 \varphi_0(\mathbf{r}) + g_2 \varphi_0^2(\mathbf{r}) .
\end{equation}

Because $V_{\text{eff}}(\mathbf{r})$ is algebraically determined by the deformation, the geometry of the potential trap directly mirrors the geometry of the soliton. If the macroscopic solitonic defect $\varphi_0(\mathbf{r})$ manifests as a localized spatial indentation, then $V_{\text{eff}}(\mathbf{r})$ acts as a deep, localized potential well.

The emergence of this static localized deformation has two direct consequences for the excitonic sector. First, because the solitonic trap is spatially localized, it breaks translational symmetry and converts the extended excitonic states of the uniform phase into a discrete set of localized wavefunctions, denoted by the $|\lambda\rangle$ basis in Fig.~3. Second, confinement inside the soliton enhances the spatial purity of these excitonic wavefunctions by suppressing their coupling to long-wavelength, low-$q$ thermal lattice fluctuations. We discuss these two consequences below.

\subsubsection*{2.2a Breaking Translational Symmetry Mixes the Extended Exciton States}

In the uniform phase, the lattice deformation is spatially constant. Because this background preserves continuous translational invariance, momentum remains a good quantum number for the unperturbed excitonic states $w_n(\mathbf{r})$, which may be represented schematically as extended plane-wave-like states in Fig.~3 of the main text. Consequently, the static interaction matrix is diagonal in momentum space, $J_{nm}\propto\delta_{nm}$, and produces only uniform energy shifts.

By contrast, the newly formed solitonic deformation $\varphi_0(\mathbf{r})$ is spatially localized. This localization breaks continuous translational invariance and relaxes momentum conservation. The localized potential therefore acts as a static scattering center, generating off-diagonal matrix elements in $J_{nm}$ that coherently mix the extended bare states $w_n(\mathbf{r})$. Because the unperturbed Hamiltonian is diagonal in its eigenbasis, the reconstructed single-particle spectrum inside the defect is described by the effective matrix $M_{nm}=E_n\delta_{nm}+J_{nm}$. Diagonalizing this matrix yields a discrete set of localized excitonic eigenstates $|\lambda\rangle$ with energies $E_\lambda$. We designate the lowest-energy state of this confined spectrum as the static ground state, $|0\rangle$.

The localized spatial wavefunctions of the solitonic trap, denoted by $w_\lambda(\mathbf{r})$, are obtained by inserting the effective potential $V_{\text{eff}}(\mathbf{r})$ into the time-independent Schr\"{o}dinger equation:

\begin{equation}
\left[ h_0 + g_1 \varphi_0(\mathbf{r}) + g_2 \varphi_0^2(\mathbf{r}) \right] w_\lambda(\mathbf{r}) = E_\lambda w_\lambda(\mathbf{r}) .
\end{equation}

Thus, one does not need to integrate over the macroscopic exciton density to define the trap itself. The confining well $V_{\text{eff}}(\mathbf{r})$ exists as a direct consequence of the spontaneous lattice deformation. Excitons in this potential organize into discrete localized levels $E_\lambda$, equivalently obtained by diagonalizing the matrix $M_{nm}=E_n\delta_{nm}+J_{nm}$.

\subsubsection*{2.2b Local Confinement Restores Spatial Purity by Suppressing Thermal Fragmentation}
In the uniform, high-temperature symmetric phase, excitons are highly vulnerable to thermal dephasing. The continuous spectrum of soft lattice vibrations (thermal acoustic phonons and low-energy optical phonons) creates a rapidly fluctuating, highly disordered potential landscape. Excitons attempting to propagate through this environment become localized by random thermal fluctuations, causing their wavefunctions to become spatially fragmented, lacking well-defined phase or structural coherence over any meaningful distance. Consequently, in the uniform phase, the bare excitonic states are highly disordered.

The solitonic transition fundamentally resolves this thermodynamic fragmentation. The localized deformation $\varphi_0(\mathbf{r})$ forms a deep, structurally rigid confining well characterized by a small microscopic volume with a characteristic length scale $L$. Because of this tight spatial confinement, the local phonon density of states inside the defect is fundamentally altered; the allowed vibrational modes become geometrically quantized. Specifically, confinement to the solitonic domain imposes a momentum infrared cutoff $q_{\text{min}} \sim \pi/L$ on the local lattice modes. For acoustic phonons with sound velocity $v_s$, this introduces a low-energy gap in the local phonon spectrum, $E_{\text{gap}} \simeq \hbar v_s \pi / L$. Because the solitonic volume is microscopically small, this local gap strictly truncates the continuous spectrum of soft, long-wavelength thermal fluctuations ($E < E_{\text{gap}}$) that typically drive dephasing. 

Therefore, excitons trapped within the soliton are no longer subjected to a noisy, disordered thermal bath. Instead, they are governed entirely by the smooth, coherent static potential of the defect. This yields sharp, discrete excitonic modes $|\lambda\rangle$ with robust, well-defined spatial coherence. This spatial purity is a critical prerequisite for the many-body physics derived in Note 3: when excitons subsequently interact and scatter within the defect via zero-mode fluctuations, they do not dissipate into an incoherent thermal continuum, but rather scatter deterministically into these spatially rigid, highly coherent excited states.

\section*{Supplementary Note 3: Dynamic Mixing of Localized Exciton Levels by the Soliton Zero Mode}

While the static eigenstates $\{|0\rangle, |\lambda\rangle\}$ capture the internal geometry of the defect, fixing the absolute position of the soliton artificially selects one point from a continuously degenerate family of translated configurations. Because the underlying medium is translationally invariant, the soliton center defines a collective coordinate associated with a translational zero mode.

Let $V_{\text{sol}}(\mathbf{r}) = g_1 \varphi_0(\mathbf{r}) + g_2 \varphi_0^2(\mathbf{r})$ represent the effective static confining potential. As the soliton freely fluctuates along the zero mode by a small spatial displacement, the potential shifts relative to the fixed spatial coordinates of the excitonic basis. Expanding the displaced potential $V_{\text{sol}}(\mathbf{r} - \delta \mathbf{R})$ to first order yields a dynamic fluctuation-induced perturbation:

\begin{equation}
\delta V(\mathbf{r}) = - \delta \mathbf{R} \cdot \nabla V_{\text{sol}}(\mathbf{r})
\end{equation}

Here, $\delta \mathbf{R}$ represents quantum fluctuations of the translational zero mode, characterized by a variance $\langle (\delta \mathbf{R})^2 \rangle$. 

This spatial mismatch explicitly mixes the newly formed localized electronic levels. Projected into the defect eigenbasis, the coupling between the localized ground state $|0\rangle$ and any excited state $|\lambda\rangle$ is governed by the matrix elements of this spatial gradient:

\begin{equation}
\mathbf{A}_{0\lambda} =
\left\langle 0 \left| \nabla V_{\text{sol}}(\mathbf{r}) \right| \lambda \right\rangle .
\end{equation}

Crucially, because the perturbation operator $\nabla V_{\text{sol}}(\mathbf{r})$ is the exact spatial derivative of the potential that originally defined the excitonic basis, its spatial profile is inherently matched to the length scales and the nodal structures of the wavefunctions $|0\rangle$ and $|\lambda\rangle$. As a result, this transition amplitude $A_{0\lambda}$ is naturally large. 

Importantly, for a single exciton, a rigid translation of the solitonic potential leaves the exact eigenvalue spectrum unchanged when all states are treated on equal footing. Nevertheless, applying second-order perturbation theory provides a valuable measure of the underlying coupling strength and the induced mixing between the localized ground state and excited states:
\begin{equation}
    \Delta E_0 = -\langle (\delta \mathbf{R})^2 \rangle \sum_{\lambda \neq 0} \frac{|\mathbf{A}_{0\lambda}|^2}{E_\lambda - E_0}
    \label{eq:single_state_zeromode_shift},
\end{equation}
where $E_0$ and $E_\lambda$ are the bare excitonic eigenenergies within the static solitonic potential. While this expression does not represent a physical lowering of the single-particle spectrum, every term in this summation is strictly negative (since $E_0 < E_\lambda$), reflecting a strong, shared mixing channel. Therefore, when extended to the many-body regime, the cooperative interaction yields a massive thermodynamic lowering. As detailed in \textbf{Equation 11 of the main text}, this collective cross-term scales as $O(N^2)$.

\subsection*{3.1 The Zero Mode Is a Slow, Collective Coordinate of the Soliton}

To appreciate the physical significance of the Eq.~\ref{eq:single_state_zeromode_shift}, one must consider the unique kinematics of the zero mode compared to standard lattice phonons, specifically regarding its effective potential and its inertial mass.

First, because the zero mode corresponds to translations of a defect in an otherwise continuous medium, it is gapless in the ideal translationally invariant limit. Displacing the defect involves no restoring force, meaning the effective spring constant is zero. Consequently, unlike gapped optical phonons whose fluctuation variances are heavily suppressed by a steep potential energy cost, the zero-mode variance $\langle (\delta \mathbf{R})^2 \rangle$ is anomalously large, constrained only by higher-order boundary conditions or collisions.

Second, the solitonic defect is a macroscopic, collective lattice deformation. Therefore, its inertial mass ($M_{\text{sol}}$) is significantly larger than the bare exciton mass. This heavy inertial mass dictates that the zero-mode dynamics occur at substantially lower frequencies than standard thermal noise. In conventional systems, fast, low-mass thermal phonons act as a random, high-frequency dephasing bath that broadens energy levels and destroys spatial coherence. In stark contrast, the heavy zero mode acts as a slow, coherent driving field. It allows the excitonic eigenstates to evolve and mix adiabatically, avoiding thermal fragmentation and ensuring that the level repulsion described above is a robust, coherent restructuring of the local Hilbert space.

\subsection*{3.2 The Soliton Displacement Couples Localized Exciton Levels}

To analytically demonstrate the scaling of the matrix elements, we evaluate a displacement along a single principal symmetry axis (e.g., x) within a 1D harmonic confining potential: 
\begin{equation}
V(x) = \frac{1}{2} m \omega^2 x^2
\end{equation}
where $m$ is the effective mass of the exciton and $\omega$ is the bare trapping frequency. In this bare potential, the energy gap between the symmetric ground state $|0\rangle$ and the first antisymmetric excited state $|1\rangle$ is exactly:
\begin{equation}
\Delta E_{\text{gap}} = \hbar \omega
\end{equation}
The oscillator length of the localized exciton is given by
$l_{\text{ex}}=\sqrt{\hbar/(m\omega)}$. The zero-mode operator, representing the gradient of this potential, is defined as:
\begin{equation}
\hat{A} = \frac{\partial V(x)}{\partial x} = m \omega^2 x
\end{equation}
We now calculate the off-diagonal transition amplitude $A_{01}$ by evaluating the matrix element:
\begin{equation}
A_{01} = \langle 0 | \hat{A} | 1 \rangle = m \omega^2 \langle 0 | x | 1 \rangle
\end{equation}
Using the standard position matrix element for a quantum harmonic oscillator, $\langle 0 | x | 1 \rangle = l_{\text{ex}} / \sqrt{2}$, we find:
\begin{equation}
A_{01} = \frac{m \omega^2 l_{\text{ex}}}{\sqrt{2}}
\end{equation}
By substituting the identity $m \omega^2 = \hbar \omega / l_{\text{ex}}^2 = \Delta E_{\text{gap}} / l_{\text{ex}}^2$, we arrive at a remarkably simple and profound scaling law for the coupling strength:
\begin{equation}
A_{01} = \frac{1}{\sqrt{2}} \frac{\Delta E_{\text{gap}}}{l_{\text{ex}}}
\end{equation}

\subsection*{3.3 Zero-Mode Mixing Is Set by the Trap Energy Scale}

This algebraic result reveals exactly why the solitonic zero mode exerts such dominant control over the single-particle states:
\begin{itemize}
    \item \textbf{Geometric lock-in:} Because the exciton wavefunction and the structural gradient are generated by the same confining potential, the coupling $A_{01}$ is not an independent small perturbative parameter. It is set by the trap energy scale divided by the trap length scale.

    \item \textbf{Non-perturbative level mixing:} If the soliton center shifts by a distance comparable to the exciton localization length, $\delta X \sim l_{\text{ex}}$, the coupling energy $A_{01}\delta X$ becomes comparable to the bare trap gap. The zero-mode displacement can therefore strongly mix the localized levels.

    \item \textbf{Many-body scaling:} In the many-body limit, this same shared displacement coordinate couples collectively to all excitons trapped in the soliton. The resulting cross-terms scale as $O(N^2)$, making the dynamically reconstructed vacuum a collective energy scale rather than a small single-particle correction.
\end{itemize}

\section*{Supplementary Note 4: Cooperative Many-Body Stabilization from the Soliton Zero Mode} 

In Supplementary Note 3, we showed that the translational zero mode provides an unusually strong coupling channel, producing large level mixing between a single localized exciton and the excited manifold via the off-diagonal transition amplitude $A_{0\lambda}$. We now extend this analysis to the many-body regime of the solitonic phase: a macroscopic many-body system containing a finite density of interacting excitons. In this collective regime, we will show that this highly efficient mixing channel becomes cooperatively enhanced, dynamically lowering the energy of the macroscopic many-body ground state.

\subsection*{4.1 The Zero Mode Provides a Single Coherent Attractive Channel}

To evaluate the collective thermodynamic state of the multi-exciton system, we must integrate out the lattice degrees of freedom to obtain an effective, exciton-only many-body Hamiltonian. Following standard field-theoretic treatments of electron-phonon coupling (e.g., the Fröhlich transformation), integrating out any generic set of lattice vibrational modes $\alpha$ yields an effective attractive interaction.

To make the origin of this interaction explicit, we consider first a single lattice mode $\alpha$ with coordinate $Q_\alpha$, stiffness $\kappa_\alpha$, and linear coupling to a generalized excitonic operator $\hat{\mathcal{A}}_\alpha$. In the static limit, the coupled Hamiltonian takes the form
\begin{equation}
H = H_{\mathrm{ex}} + \frac{\kappa_\alpha}{2} Q_\alpha^2 - Q_\alpha \hat{\mathcal{A}}_\alpha .
\end{equation}
Completing the square in $Q_\alpha$ yields
\begin{equation}
H =
H_{\mathrm{ex}}
+ \frac{\kappa_\alpha}{2}
\left(
Q_\alpha - \frac{\hat{\mathcal{A}}_\alpha}{\kappa_\alpha}
\right)^2
- \frac{1}{2\kappa_\alpha} \hat{\mathcal{A}}_\alpha^2 .
\end{equation}
Since the first term is positive definite, integrating out the lattice coordinate $Q_\alpha$ leaves an effective exciton-only contribution
\begin{equation}
H_{\mathrm{eff}}^{(\alpha)} = H_{\mathrm{ex}} - \frac{1}{2\kappa_\alpha} \hat{\mathcal{A}}_\alpha^2 .
\end{equation}
For a set of independent lattice modes $\alpha$, the total effective interaction is obtained by summing over all modes, yielding
\begin{equation}
H_{\text{eff}} = H_{\text{ex}} - \sum_\alpha \frac{1}{2\kappa_\alpha} \hat{\mathcal{A}}_\alpha^2 .
\end{equation}
Here $H_{\text{ex}}$ is the bare many-body Hamiltonian, and $\kappa_\alpha$ represents the effective dynamic stiffness (energetic restoring force) of the corresponding lattice mode.

For standard vibrational modes, including the broad continuum of bulk phonons in the homogeneous crystal as well as the gapped internal shape and breathing modes localized within the solitonic defect, the stiffness $\kappa_\alpha$ is finite and typically large. Consequently, the resulting attractive interactions ($\sim 1/\kappa_\alpha$) are weak. Moreover, because these interactions are distributed over a broad manifold of distinct modes $\alpha$, the sum over this large phase space produces incoherent, competing fluctuations. Rather than condensing into a single collective channel, these generic modes act as an effective thermal bath that disrupts phase coherence and suppresses off-diagonal long-range order (ODLRO).

In contrast, among all lattice degrees of freedom, the continuous sliding of the defect along its translational zero mode ($\alpha = 0$) is fundamentally unique. As established in Note 3, this mode arises from spontaneous breaking of continuous translational symmetry and is gapless in the ideal limit. In a physical system, its restoring force is only weakly lifted by higher-order processes (e.g., collisions, finite-size effects, or inertial constraints), resulting in an anomalously small effective stiffness $\kappa_0 \equiv \kappa$.

Because this zero-mode deformation acts globally over the confined volume, its associated structural gradient couples uniformly to all excitons within the potential. For clarity, we first consider a zero-mode displacement along a single direction, denoted by $X$. The corresponding collective coupling operator is
\begin{equation}
\hat{\mathcal{A}}_X =
\sum_{i=1}^{N}
\hat{\mathbf e}_X \cdot \nabla_i V_{\text{sol}}(\mathbf{r}_i) .
\end{equation}
The effective interaction mediated by this zero-mode coordinate is then
\begin{equation}
    H_{\text{eff}} = H_{\text{ex}} - \frac{1}{2\kappa_X} \hat{\mathcal{A}}_X^2 .
\end{equation}
In the following, we suppress the subscript $X$ and write $\hat{\mathcal{A}}$ and $\kappa$ for this chosen zero-mode direction.

Because this interaction is mediated by a single collective coordinate shared by all excitons in the solitonic domain, it acts as an all-to-all coherent attractive channel within that domain. The remaining lattice modes are not eliminated physically; they remain present as incoherent thermal and dephasing channels. In the effective theory, they are treated as a competing thermal background, while the zero mode is isolated as the dominant coherent contribution. The ordered state is stable when the cooperative zero-mode energy lowering exceeds the thermal and incoherent fluctuations generated by the broader phonon bath.

\subsubsection*{Validity of the static treatment of the zero mode}
We treat the zero-mode coordinate as a static background and solve for the lowest-energy self-consistent exciton configuration, rather than for a dynamical, retarded interaction. This is justified by the defining feature of the solitonic phase: all relevant excitons are bound within the \emph{same} confining trap and couple to a \emph{single} shared coordinate, i.e., the position of that trap. In contrast to a conventional phonon-mediated interaction, in which one excitation perturbs the lattice and the disturbance must \emph{propagate} across a finite distance to reach a second, spatially separated excitation (so that the mediated interaction is inherently delayed and frequency-dependent, as in the retarded electron--phonon coupling of BCS), the co-located excitons here all respond to the same rigid displacement of their common trap at the same instant. There is no propagation distance and hence no travel time, so the interaction carries no retardation. Equivalently, the soft translational zero mode ($\hbar\omega_0 \to 0$) is slow compared with the finite internal excitonic level spacing $E_{10}$ set by the confining potential, so the fast internal states track the slow coordinate adiabatically in the Born--Oppenheimer sense. Evaluating the self-consistent static minimum by completing the square is therefore exact, and retardation effects do not enter.

\subsection*{4.2 The Cooperative Energy Lowering Scales as \(N^2\)}

To evaluate the thermodynamic ground state of this zero-mode-mediated attractive interaction, we expand the collective zero-mode operator around its macroscopic expectation value. Defining $\hat{\mathcal{A}} = \langle \hat{\mathcal{A}} \rangle + \delta\hat{\mathcal{A}}$ and neglecting the second-order fluctuations ($\delta\hat{\mathcal{A}}^2 \approx 0$) at the mean-field level, the squared interaction term expands as:
\begin{equation}
\hat{\mathcal{A}}^2 \approx \langle \hat{\mathcal{A}} \rangle^2 + 2\langle \hat{\mathcal{A}} \rangle (\hat{\mathcal{A}} - \langle \hat{\mathcal{A}} \rangle) = 2\langle \hat{\mathcal{A}} \rangle \hat{\mathcal{A}} - \langle \hat{\mathcal{A}} \rangle^2
\end{equation}

To describe the stabilizing effect of a finite collective distortion field, we define the self-consistent macroscopic driving field as $\Delta = \frac{\langle \hat{\mathcal{A}} \rangle}{\kappa}$. Substituting this definition back into the expanded interaction yields the mean-field many-body Hamiltonian:
\begin{equation}
H_{\text{MF}} = H_{\text{ex}} - \Delta \hat{\mathcal{A}} + \frac{1}{2} \kappa \Delta^2
\end{equation}

This expression captures the thermodynamic competition fundamental to the phase transition. The operator $-\Delta \hat{\mathcal{A}}$ acts as a self-generated driving field that mixes the bare states to lower the collective electronic energy, while the positive scalar term $\frac{1}{2} \kappa \Delta^2$ represents the macroscopic elastic penalty paid by the lattice to sustain the continuous structural deformation.

To find the total collective ground-state energy of the system ($E_{\text{MF}}$), we take the expectation value of this entire mean-field Hamiltonian:
\begin{equation}
E_{\text{MF}} = \langle H_{\text{MF}} \rangle = E_{\text{bare}} - \Delta \langle \hat{\mathcal{A}} \rangle + \frac{1}{2} \kappa \Delta^2
\end{equation}

where $E_{\text{bare}}=\langle H_{\text{ex}}\rangle$ denotes the excitonic contribution evaluated in the corresponding mean-field state.

Because $\langle \hat{\mathcal{A}} \rangle = \kappa \Delta$ by our definition, the total mean-field energy simplifies to:
\begin{equation}
E_{\text{MF}} = E_{\text{bare}} - \frac{1}{2} \kappa \Delta^2
\end{equation}

This result provides a mean-field demonstration of the dynamic vacuum reconstruction. If the system remains in the unperturbed, symmetric state, the macroscopic field is zero ($\Delta = 0$), and the energy remains at the bare excitonic level. However, if the excitons dynamically hybridize into a state with a finite collective driving field, the total energy of the system is strictly reduced. 

This same elastic term resolves a concern intrinsic to soft modes. Although the bare translational mode is soft, the self-consistent state is not: the positive $\tfrac{1}{2}\kappa\Delta^2$ contribution gives the collective coordinate a finite restoring curvature about the ordered configuration, with an energy cost for displacement set by the macroscopic stabilization scale $\sim N^2 A_{0\lambda}^2/2\kappa$. The soft mode that enables the ordered state is therefore stiffened by the order it creates, so its thermal fluctuations do not wash out the coherence.

As established in Note 3, because the perturbation operator is geometrically matched to the structural length scale of the solitonic potential, the single-particle transition amplitude $A_{0\lambda}$ is naturally large. For a macroscopic density of $N$ localized excitons occupying the same solitonic volume, the collective expectation value scales as $\langle \hat{\mathcal{A}} \rangle \sim N A_{0\lambda}$. Consequently, the macroscopic driving field scales directly with the density:
\begin{equation}
\Delta \approx \frac{N A_{0\lambda}}{\kappa}
\end{equation}
Plugging this dynamically generated field into our total energy equation reveals that the collective many-body ground state is shifted downward by:
\begin{equation}
\Delta E_{\text{collective}} = - \frac{1}{2} \kappa \Delta^2 \approx - \frac{N^2 A_{0\lambda}^2}{2\kappa}
\end{equation}

Because this collective energy lowering scales as $N^2$, the stabilization energy per exciton grows linearly with exciton number within the solitonic volume. This cooperative enhancement turns the zero-mode coupling into a density-dependent protection energy, allowing the reconstructed many-body state to compete with thermal fluctuations.

\section*{Supplementary Note 5: Suppression of Thermal Phase Space and Enhancement of the Critical Temperature}

To quantitatively demonstrate the thermodynamic stability of the macroscopic quantum state, we evaluate the phase-space capacity of the excitonic system within the fluctuation-renormalized spectrum. The total number of excitons is conserved and partitioned between the macroscopic collective mode ($N_0$) and the excited-state thermal continuum ($N_{\text{ex}}$):
\begin{equation}
    N = N_0 + N_{\text{ex}}(T)
\end{equation}
At the critical temperature $T = T_c$, the macroscopic population of the collective mode just begins to form ($N_0 \to 0$), meaning the thermal continuum is entirely saturated: $N = N_{\text{ex}}(T_c)$.

\subsection*{5.1 The Zero-Mode Energy Lowering Pins the Chemical Potential Below the Continuum}
The excited-state population may contain contributions from both discrete bound levels and the continuum above the confinement threshold. Here, for analytical transparency, we approximate the thermally accessible excited manifold by an effective continuum with lower edge $E_c$.  This treatment is conservative, because it overcounts the available excited-state phase space and therefore underestimates the resulting enhancement of macroscopic occupation. The standard density of states for 3D is $g(E) = \frac{V}{4\pi^2} \left( \frac{2m^*}{\hbar^2} \right)^{3/2} \sqrt{E - E_c}$ for $E \ge E_c$.

In a standard, translationally invariant vacuum, the ground state sits precisely at the continuum edge ($E_c$). However, as established in previous sections, coupling to the solitonic zero mode lowers the collective ground state to a renormalized energy $\tilde{E}_0$.  This dynamically generates a density-dependent stabilization gap:
\begin{equation}
    \Delta_{\text{gap}} = E_c - \tilde{E}_0 .
\end{equation}
At the onset of macroscopic condensation ($T_c$), the chemical potential $\mu$ must be pinned to the lowest available energy state. Therefore, in our dynamically reconstructed vacuum, the chemical potential is pinned below the continuum at $\mu = \tilde{E}_0$.

The total population of the saturated excited-state continuum is given by integrating the Bose--Einstein distribution over the density of states:
\begin{equation}
    N = \int_{E_c}^{\infty} \frac{g(E)}{\exp\left(\frac{E - \tilde{E}_0}{k_B T_c}\right) - 1} dE
\end{equation}

\subsection*{5.2 The Gap Reduces the Thermal Capacity of the Excited States}
To evaluate this integral, we express the exponent in terms of the collective gap $\Delta_{\text{gap}}$. Noting that $E - \tilde{E}_0 = (E - E_c) + \Delta_{\text{gap}}$, we introduce the dimensionless energy variable $x = \frac{E - E_c}{k_B T_c}$. 

The integration limits shift from $[E_c, \infty)$ to $[0, \infty)$, and the exponent strictly factorizes, allowing us to extract the gap-dependent term as an effective fugacity, $z$:
\begin{equation}
    z = \exp\left(-\frac{\Delta_{\text{gap}}}{k_B T_c}\right)
\end{equation}
Substituting this into the population integral yields:
\begin{equation}
    N = \frac{V}{4\pi^2} \left( \frac{2m^* k_B T_c}{\hbar^2} \right)^{3/2} \int_0^{\infty} \frac{\sqrt{x}}{z^{-1}e^x - 1} dx
\end{equation}
This standard integral is analytically evaluated using the polylogarithm (Bose function) of order $3/2$, defined as $g_{3/2}(z) = \frac{1}{\Gamma(3/2)} \int_0^\infty \frac{x^{1/2}}{z^{-1}e^x - 1}dx$. Using $\Gamma(3/2) = \sqrt{\pi}/2$, we obtain the exact equation for the phase-space capacity:
\begin{equation}
    n = \left( \frac{m^* k_B T_c}{2\pi\hbar^2} \right)^{3/2} g_{3/2}\left(e^{-\Delta_{\text{gap}}/k_B T_c}\right)
    \label{eq:phase_space_capacity}
\end{equation}
where $n = N/V$ is the exciton density. 

\subsection*{5.3 Phase-Space Collapse Enhances the Critical Temperature}
Equation~\ref{eq:phase_space_capacity} reveals the central thermodynamic consequences of the zero-mode mechanism. The function $g_{3/2}\left(e^{-\Delta_{\text{gap}}/k_B T_c}\right)$ directly quantifies the maximum number of thermally accessible states in the continuum. 

In a conventional uniform lattice ($\Delta_{\text{gap}} = 0$), the fugacity is exactly $1$, and the Bose function reduces to a fundamental maximum bounded by the Riemann zeta function: $g_{3/2}(1) = \zeta(3/2) \approx 2.612$. However, because zero-mode coupling generates a finite $\Delta_{\text{gap}} > 0$, the fugacity is reduced below unity and becomes exponentially small when $\Delta_{\text{gap}} \gg k_B T_c$.

As explicitly plotted in the main text, as the dimensionless gap $\Delta_{\text{gap}}/k_B T_c$ increases, the thermal capacity $g_{3/2}(z)$ rapidly collapses toward zero. This mathematical collapse provides the physical basis for the Quantum Analog of Vibration Isolation (QAVI): the dynamically generated gap suppresses the thermally accessible phase space of the excited-state manifold, so additional excitons preferentially accumulate in the collective ground state.

To determine the critical temperature enhancement, we compare Equation~\ref{eq:phase_space_capacity} to the critical temperature of a standard, static condensate ($T_c^{(0)}$) at the identical density $n$, which is bounded by $\zeta(3/2)$. Rearranging for the ratio of the temperatures directly yields (Eq. 11 in the main text):
\begin{equation}
    \frac{T_c}{T_c^{(0)}} = \left[ \frac{\zeta(3/2)}{g_{3/2}\left(e^{-\Delta_{\text{gap}}/k_B T_c}\right)} \right]^{2/3}
\end{equation}

Because the denominator $g_{3/2}$ decreases as the fugacity is reduced, the ratio $T_c / T_c^{(0)}$ increases. Furthermore, because the gap itself scales cooperatively with density, $\Delta_{\text{gap}} \propto n$, the reduction of $g_{3/2}$ creates a nonlinear feedback between density, phase-space suppression, and macroscopic occupation. This QAVI-enhanced feedback raises the transition temperature beyond the standard algebraic scaling of an unrenormalized spectrum, allowing macroscopic occupation to persist at temperatures higher than expected for the original excitation landscape.

\section*{Supplementary Note 6: Emergence of ODLRO from the Low-Rank Zero-Mode Interaction}

In Supplementary Note 4, we established that the anomalously compliant translational zero mode projects the many-body interaction into a single dominant macroscopic channel, yielding the effective Hamiltonian $H_{\text{eff}} = H_{\text{ex}} - \frac{1}{2\kappa}\hat{\mathcal{A}}^2$. In Supplementary Note 5, we then showed that the resulting zero-mode energy lowering reduces the thermal capacity of the excited-state manifold, so that increasing excitation density increasingly favors occupation of the collective ground state. Here, we examine the algebraic structure of the squared collective operator and show that the all-to-all coupling mediated by the zero mode selects a hybridized exciton many-body ground-state orbital. We then demonstrate that macroscopic occupation of this orbital satisfies the Penrose--Onsager criterion for off-diagonal long-range order (ODLRO).

\subsection*{6.1 The Shared Zero Mode Creates an All-to-All Separable Interaction}

To understand how this projected zero-mode interaction favors macroscopic occupation of a single orbital, we expand the algebraic structure of the squared collective operator. The zero-mode coupling is the sum of the same projected spatial-gradient operator acting on each of the $N$ confined excitons:
\begin{equation}
\hat{\mathcal{A}} = \sum_{i=1}^N \hat{A}_i,
\quad
\hat{A}_i = \hat{\mathbf e}_X \cdot \nabla_i V_{\text{sol}}(\mathbf{r}_i) .
\end{equation}
Integrating out this mode yields an effective many-body interaction term that is simply the square of this collective coordinate:
\begin{equation}
    -\frac{1}{2\kappa} \hat{\mathcal{A}}^2 = -\frac{1}{2\kappa} \left( \sum_{i=1}^N \hat{A}_i \right)^2 = -\frac{1}{2\kappa} \sum_{i, j} \hat{A}_i \hat{A}_j.
\end{equation}

This expansion reveals the key algebraic structure of the zero-mode mechanism. Generic two-particle interactions, such as screened Coulomb repulsion or standard bulk phonon scattering, depend intricately on the relative distance between particles, $V(\mathbf{x_i} - \mathbf{x_j})$. When transformed into a generic basis, such interactions yield "full-rank" interaction matrices that scatter excitons across a broad manifold of momentum and internal states. 

By contrast, the zero-mode interaction kernel is strictly separable. The effective coupling between any pair of excitons $i$ and $j$ does not depend on their relative coordinate; it is strictly the product of their individual interactions with the global defect gradient, $\hat{A}_i \hat{A}_j$. Because this interaction matrix is generated entirely by the outer product of a single one-body operator with itself ($\hat{A} \otimes \hat{A}$), it has a separable, rank-one structure in the zero-mode coupling channel. 

\subsection*{6.2 The Cooperative Interaction Selects a Hybridized Exciton Orbital}

To explicitly determine the structure of the dynamically reconstructed vacuum, we use the Variational Principle to minimize the total many-body energy. Based on the parity selection rules established in Note 3, the spatial gradient operator couples the even-parity ground state $\ket{0}$ to the first odd-parity excited state $\ket{1}$. We denote the bare energy spacing between these two discrete solitonic levels as $E_{10} = E_1 - E_0$. 

We use a variational ansatz in which the excitons occupy a single hybridized single-particle orbital with real amplitudes $u$ and $v$:
\begin{equation}
    \ket{\Phi_0} = u\ket{0} + v\ket{1}, \quad \text{where} \quad u^2 + v^2 = 1
\end{equation}
For an $N$-particle trial state $\ket{\Psi_N} = \bigotimes_{i=1}^N \ket{\Phi_0}_i$, the bare energy cost scales linearly: $E_{bare} = N v^2 E_{10}$.

The collective interaction energy is generated by the squared operator $\hat{\mathcal{A}}^2 = (\sum_{i=1}^N \hat{A}_i)^2$. Expanding this yields $N$ self-energy terms and $N(N-1)$ cooperative cross-terms ($\hat{A}_i\hat{A}_j$). The expectation value of a single operator $\hat{A}$ in the trial state is $\bra{\Phi_0} \hat{A} \ket{\Phi_0} = 2uv A_{01}$. Consequently, the cooperative cross-term interaction energy is:
\begin{equation}
    E_{int, cross} = -\frac{1}{2\kappa} \Big( N(N-1) (2uv A_{01})^2 \Big) = - \frac{2 N(N-1) A_{01}^2}{\kappa} u^2 v^2.
\end{equation}

Formulating the total energy $E_{total} = E_{bare} + E_{int, cross}$ and substituting $x = v^2$ (using $u^2 = 1 - x$), we obtain:
\begin{equation}
    E_{total}(x) = Nx E_{10} - \frac{2 N(N-1) A_{01}^2}{\kappa} x(1-x) + \mathcal{O}(N).
\end{equation}
To determine the optimal mixing fraction, we minimize this energy with respect to $x$:
\begin{equation}
    \frac{dE_{total}}{dx} = N E_{10} - \frac{2 N(N-1) A_{01}^2}{\kappa} (1 - 2x) = 0.
\end{equation}
Solving for $x$ yields the exact density-dependent mixing parameter $v_N^2$:
\begin{equation}
    v_N^2 = \frac{1}{2} \left( 1 - \frac{E_{10} \kappa}{2 (N-1) A_{01}^2} \right).
\end{equation}

This variational result illustrates the nonlinear feedback produced by the zero-mode interaction. As the particle number $N$ increases, the cooperative interaction increasingly overcomes the single-particle level-spacing cost. Within this two-level variational model, the macroscopic limit gives an equal-weight superposition:
\begin{equation}
    \lim_{N \to \infty} v_N^2 = \frac{1}{2} \implies u = v = \frac{1}{\sqrt{2}}.
\end{equation}
This shows that, in the macroscopic limit, the cooperative zero-mode energy can overcome the bare excitonic level spacing and drive the system toward a strongly hybridized collective orbital. 

\subsection*{6.3 Macroscopic Occupation Satisfies the Penrose--Onsager Criterion}

The essential consequence of the zero-mode interaction is that it is separable. In the localized solitonic basis $\{\ket{\lambda}\}$, the translational zero mode introduces the coupling matrix:
\begin{equation}
A_{\lambda\lambda'} =
\bra{\lambda}\hat{\mathbf e}_X\cdot\nabla V_{\text{sol}}\ket{\lambda'}.
\end{equation}
Because the effective interaction is generated by the square of a single collective operator, $H_{int}^{(0)} = -\frac{1}{2\kappa}\hat{A}^2$, it does not represent a generic full-rank scattering interaction. Instead, it projects the many-body dynamics into one dominant collective channel.

The corresponding single-particle orbital is a zero-mode-dressed superposition of the localized solitonic states:
\begin{equation}
    \ket{\Phi_0} = \sum_\lambda f_\lambda \ket{\lambda}, \quad b^\dagger_{\Phi_0} = \sum_\lambda f_\lambda b^\dagger_\lambda .
\end{equation}
The coefficients $f_\lambda$ are determined by minimizing the full effective Hamiltonian, including both the static localized spectrum and the zero-mode-mediated interaction. 

While this orbital is formally defined as a general superposition over the entire localized spectrum, the physical distribution of the coefficients $f_\lambda$ is strictly governed by parity selection rules. As established in Supplementary Note 3, the zero-mode operator $\hat{A}$, being proportional to $\nabla V_{\text{sol}}$, has odd parity. Consequently, its matrix elements connect states of opposite parity, with the dominant coupling occurring between the even-parity ground state $\ket{0}$ and the first odd-parity excited state $\ket{1}$. Coupling to same-parity states, such as $\bra{0}\hat{A}\ket{2}$, is forbidden by parity. Therefore, when the lowest opposite-parity excitation dominates the matrix element, the condensed orbital is well approximated by the primary two-level subspace:
\begin{equation}
    \ket{\Phi_0} \approx f_0\ket{0} + f_1\ket{1} \equiv u\ket{0} + v\ket{1}.
\end{equation}
In the strong zero-mode-coupling regime, the dominant structure of $\ket{\Phi_0}$ is controlled by the leading eigenvector of this truncated channel.

To see why this orbital becomes macroscopically occupied, consider the $N$-exciton trial state:
\begin{equation}
    \ket{\Psi_N} = \frac{1}{\sqrt{N!}} (b^\dagger_{\Phi_0})^N \ket{vac}.
\end{equation}
For this state, the expectation value of the collective operator scales as $\langle\hat{A}\rangle = N\bra{\Phi_0}\hat{A}\ket{\Phi_0}$. Consequently, the zero-mode contribution to the energy contains a coherent collective term:
\begin{equation}
    \langle H_{int}^{(0)} \rangle \simeq -\frac{N(N - 1)}{2\kappa} \Big| \bra{\Phi_0}\hat{A}\ket{\Phi_0} \Big|^2 + \mathcal{O}(N).
\end{equation}
This term scales as $\mathcal{O}(N^2)$ and therefore dominates over ordinary single-particle self-energy corrections, which scale only as $\mathcal{O}(N)$. Excitons occupying orbitals orthogonal to the selected collective channel do not acquire the same coherent $\mathcal{O}(N^2)$ stabilization. The thermodynamic ground state is therefore energetically biased toward macroscopic occupation of the reconstructed orbital $\ket{\Phi_0}$.

The one-body density matrix in the localized basis is:
\begin{equation}
    \rho_{\lambda\lambda'} = \bra{\Psi_N} b^\dagger_\lambda b_{\lambda'} \ket{\Psi_N} = N f^*_\lambda f_{\lambda'}.
\end{equation}
This matrix is rank one for the condensed component and has a leading eigenvalue:
\begin{equation}
    \lambda_0 = N
\end{equation}
with all orthogonal eigenvalues vanishing in the ideal condensed limit. In the presence of thermal or non-condensed excitations, this result becomes:
\begin{equation}
    \lambda_0 = \mathcal{O}(N), \quad \lambda_{\mu\neq0} = \mathcal{O}(1),
\end{equation}
which is the Penrose--Onsager criterion for off-diagonal long-range order.

\end{document}